# Mid-infrared Assisted THz Phonon Amplification in a 2D Semiconductor for Room Temperature Detection

Christopher Sumner[1], Jakob Ziewer[2], Anju Sajan[1], Fumin Huang[2] and Rohit Chikkaraddy[1*]

[1]School of Physics and Astronomy, University of Birmingham, Birmingham, B15 2TT, UK

[2]School of Mathematics and Physics, Queens University Belfast, Belfast, BT7 1NN, UK

ABSTRACT.

Efficient and selective excitation of lattice vibrations is central to controlling energy flow at the nanoscale, yet remains challenging under conventional optical excitation. Here we introduce a mid-infrared assisted phonon amplification (MIRAPA) approach that enables efficient energy injection directly into vibrational bonds. Using surface-enhanced resonant Raman scattering in few-layer $MoS_2$, we exploit strong exciton-phonon coupling to monitor phonon populations. When mid-infrared (MIR) light is introduced, it couples directly to the out-of-plane lattice vibrations, leading to a room temperature phonon amplification exceeding 80%. Crucially, MIRAPA bypasses electronic excitation pathways, allowing the MIR power density to be nearly 300× lower than that required for visible excitation to achieve comparable enhancement. The resulting phonon modulation is robust persisting over more than 2800 on/off cycles and exceeding 15 hours of continuous wave laser illumination without degradation. Quantitative analysis yields an effective noise-equivalent power of $\sim 0.3 \mathrm{nW}/\sqrt{\mathrm{Hz}}$ for MIR detection, highlighting the sensitivity of the approach. By combining vibrational selectivity, low-power operation, and long-term stability, MIRAPA provides a robust platform for probing and amplifying phonons in two-dimensional semiconductors. These results open new opportunities for nanoscale vibrational sensing, mid-infrared detection, and phonon-based coherent devices, including routes towards phonon lasing.

## Introduction

Inelastic light scattering provides a unique methodology to convert an optical field into vibrational excitations in quantum materials[1,2]. This offers an optical route to feed energy into, or extract energy from, quantum systems, particularly for low energy excitations at terahertz (THz) frequencies[3,4]. Such control enables advances in light-controlled heat transport, superconductivity, lasing, and highly efficient sensors[5–10]. Transition metal dichalcogenides (TMDs) are interesting platforms for such studies due to strong visible excitonic resonances, robust THz phonon modes, and substantial electron-phonon coupling[11–17], all of which are crucial for efficient energy conversion and interaction.

In TMDs, the high exciton quantum efficiency is expected owing to the large exciton binding energy[17–19]. The resulting enhanced light-matter interaction has enabled sideband Raman cooling[20–22], enhanced correlated Stokes and anti-Stokes scattering[23,24], and phonon-mediated upconversion processes[25–28]. However, efficient and mode-selective phonon amplification or cooling is often hindered by defects, structural inhomogeneities, and non-radiative recombination pathways[29–33]. Even modest non-radiative optical absorption can disrupt coherence, especially because the pump photon energy is typically an order of magnitude larger than the THz phonon resonance.

To overcome these limitations, we propose an optomechanical pumping strategy, in which phonon system is MIR-primed (Fig.1c). By tuning the mid-infrared (MIR) photon energy close to THz phonon frequencies, this approach provides a near-resonant coupling to phonon modes. Unlike visible or near-infrared excitation which can introduce excessive heating and decoherence, MIR excitation can prime the phonon system while minimising electronic absorption. Here we demonstrate this concept in air stable molybdenum disulphide ($MoS_2$) multilayer stacks: MIR priming produces a substantial increase in out-of-plane THz phonon population with minimal effects on in-plane phonons. We further show that MIR-assisted phonon amplification (MIRAPA) provides a route towards MIR upconversion detection schemes relevant to biological, astronomical and industrial sensing[34–39].

## MIR-assisted phonon amplification in $MoS_2$

Surface enhanced resonant Raman scattering (SERRS) spectra of $MoS_2$ are measured using a custom built dual-channel microscope combining ultra-narrow volume holographic grating (VHG) notch filters with a liquid-nitrogen cryostat (Fig.1d). This enables simultaneous detection of Stokes ($S$) and anti-Stokes ($aS$) Raman scattering down to ±10cm$^{-1}$ Raman shifts, together with white-light reflection spectra of $MoS_2$ across temperatures of 105K-300K. We use $E_L$=1.95eV ($\lambda_L$=633nm) to resonantly pump close to $\varepsilon_\mathrm{A}$ and $\varepsilon_\mathrm{B}$ excitons in few-layer $MoS_2$ (~5nm) exfoliated on a 15nm thick gold (Au) foil (Methods). The finite thickness Au substrate provides both front- and back-side access to $MoS_2$ (a 2D-excitons-on-foil geometry, 2DoF), allowing MIR pumping from the top side (Fig.1d) while the visible laser incident through the glass side. The 633nm excitation is in resonance with the $\varepsilon_\mathrm{A}$ and $\varepsilon_\mathrm{B}$ excitons attributed to transitions between the spin-orbit split valence band and the lowest conduction band at the **K** (or **K'**) point in the Brillouin zone[16,17] (Fig.1b).

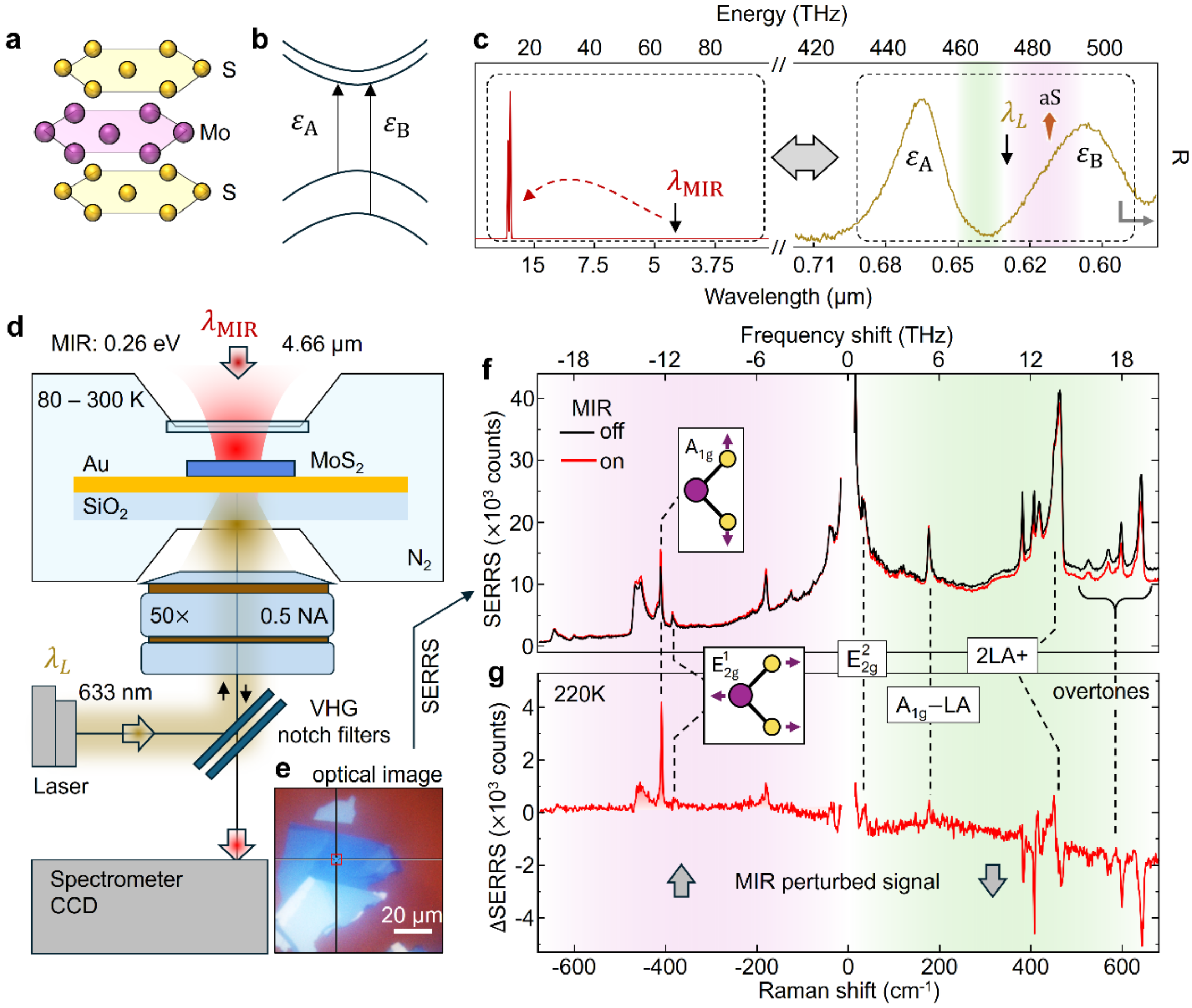

**Figure 1: MIR excitation and THz-SERRS of $MoS_2$ phonons.** (**a**) Schematic of a monolayer $MoS_2$ crystal showing the trigonal prismatic coordination of Mo atoms (purple) sandwiched between two layers of S atoms (yellow). (**b**) Energy-level diagram illustrating optical transitions at the **K**/**K′** valleys, giving rise to the A and B excitonic resonances ($\varepsilon_A$ and $\varepsilon_B$). (**c**) Spectral schematic linking mid-infrared (MIR) excitation to visible optical readout. Left: MIR excitation at wavelength $\lambda_{\text{MIR}}$ in the THz regime with the Raman active peaks indicated. Right: Reflectance ($R$) spectrum at 105K normalised to the background reflectance from the gold substrate in the visible regime showing the $\varepsilon_A$ and $\varepsilon_B$ resonances, with SERRS probing at wavelength $\lambda_L$ and detection of anti-Stokes (aS) scattering. (**d**) Experimental schematic of the MIR-pumped SERRS setup. A focused $\lambda_{\text{MIR}}$ illuminates $MoS_2$ supported on an Au/$SiO_2$ substrate over the temperature range 80–300 K. $\lambda_L$ is coupled through a high-NA objective, with back-scattered light filtered by volume holographic notch filters and detected using a spectrometer and CCD. (**e**) Optical image of a $MoS_2$ flake with crosshair representing excitation region. (**f**) Stokes and anti-Stokes SERRS spectrum with (red) and without (black) MIR, showing phonon modes including $\text{A}_{1g}$, $\text{E}^1_{2g}$, 2LA combinations, and overtones. (**g**) Difference in SERRS spectra in **f**, isolating MIR-induced modulation of the phonon populations.

The observed SERRS comprises first-order Raman-active vibrational modes of $MoS_2$ at the $\Gamma$ point of the Brillouin zone[40–44]: $\text{E}^2_{2g}$, $\text{E}^1_{2g}$ and $\text{A}_{1g}$ at 35cm$^{-1}$, 380cm$^{-1}$ and 408cm$^{-1}$, respectively, assigned according to their polarization properties (Fig.1f). Resonant pumping at $\lambda_L$=633nm produces additional peaks due to higher-order Raman scattering, enhanced by coupling of phonon modes to electronic transitions. The most prominent is the 2LA peak, together with its overtones[44–47,14]. The broad feature at 460cm$^{-1}$ labelled 2LA+, comprises multiple contributions and can be decomposed into least three peaks[14,47,48]. While the dispersion and ordering of these modes remain debated[14,42,46,48,49], there is general agreement that the higher-energy contributions arise from two longitudinal acoustic phonons (2LA) from the **K** and **M** points in the Brillouin zone. Combinations with other modes are also possible, and we observe an $\mathrm{A_{1g}} - \mathrm{LA}$ peak at 179cm$^{-1}$ (Fig.1f)[49]. The $aS$ side of the spectrum mirrors these peaks with different relative intensities.

In the presence of MIR illumination at $\lambda_{\mathrm{MIR}}$= 4.65μm (267meV) the $aS$ side of the SERRS spectrum shows pronounced increase in the intensities of $\mathrm{A_{1g}}$ , $\mathrm{A_{1g}}$ – LA and 2LA+ modes, whereas $\mathrm{E_{2g}^2}$, $\mathrm{E_{2g}^1}$ and background features remain weakly affected (Fig.1g). In contrast, on the $S$ side both $\mathrm{A_{1g}}$ and $\mathrm{E_{2g}^1}$ decrease by ~10%, indicating that the MIR response is not a uniform intensity scaling. The out-of-plane $\mathrm{A_{1g}}$ phonon exhibits a stronger response to electronic coupling due to orbital overlap between excitons and lattice motion in $MoS_2$. The $d_{z^2}$ orbitals of the molybdenum atom contribute strongly to $\varepsilon_A$ and $\varepsilon_B$ exciton formation near 600–660 nm, producing resonances close to the excitation used here[16,50]. Consequently, out-of-plane electronic motion couples more strongly to out-of-plane atomic motion, enhancing coupling to the $\mathrm{A_{1g}}$ phonon. This is further enhanced by the image charge coupling[51] to Au foil.

The selective enhancement of $aS$ scattering cannot be explained by excitonic bleaching alone, because bleaching would reduce both $S$ and $aS$ intensities similarly. Instead, the data indicates a MIR-induced increase in phonon occupation. Conceptually, this can be viewed as a MIR-assisted Raman process in which MIR field couples to out-of-plane phonons, effectively priming the lattice prior to the 633nm Raman probe. The MIR-assisted $aS$ signal for the out-of-plane $\mathrm{A_{1g}}$ phonon increases by >50%, whereas the in-plane $\mathrm{E_{2g}^1}$ mode changes by <2%, demonstrating mode selective amplification. As a result, the phonon population $\bar{n}$ of the $\mathrm{A_{1g}}$ mode (at $\Omega_v$=12.2THz) is driven above the thermal population expected at the equilibrium temperature $T$ of the environment,

$$\bar{n}_T = \left[\exp\left(\frac{\hbar\Omega_v}{k_B T}\right) - 1\right]^{-1} \quad (1)$$

satisfying the Bose−Einstein distribution, where $\hbar$ is the reduced Planck constant and $k_B$ is the Boltzmann constant.

The MIR-induced change in phonon population is estimated from the $aS$ to $S$ intensity ratio. Under thermal equilibrium the ratio $aS$ / $S$ scales as $\bar{n}_T/(\bar{n}_T + 1)$. For the $\mathrm{A_{1g}}$ phonon at a sample temperature ($s_T$) of 200K the expected $aS/S$ ratio is 0.056. Experimentally, however, we measure a ratio of 0.58 without MIR and 1.16 with MIRAPA representing an order-of-magnitude deviation from the thermal expectation. Such large departures from equilibrium can occur under near-exciton pumping due to resonance effects that selectively enhance or suppress Stokes or anti-Stokes photons via virtual electron–hole pairs, correlated Stokes–anti-Stokes scattering processes in which a Stokes photon is effectively consumed during an anti-Stokes event sharing the same vibrational quantum, and exciton-mediated optomechanical coupling[3,24,52–56]. As we show below, under MIR-assisted SERRS the $aS$ intensity scales with MIR power and suppression in linewidths, consistent with exciton-mediated optomechanical coupling as the dominant contribution. We also performed experiments at $\lambda_L$=785nm (SI1), detuned from the exciton resonance, and do not observe MIR-assisted modulation in the SERS signal. Thus, further supporting the role of resonant electron-phonon coupling.

In the regime of exciton resonance mediated optomechanical coupling the ratio of $aS$ to $S$ scattering is then,

$$\frac{I_{as}}{I_s} \approx \left(\frac{\omega_L + \Omega_v}{\omega_L - \Omega_v}\right)^4 \left(\frac{\mathcal{L}_-}{\mathcal{L}_+}\right) \zeta \left(\frac{\bar{n}}{1+\bar{n}}\right) \tag{2}$$

where $\bar{n}$ is the phonon occupation number, $\zeta$ is the wavelength dependent emission and detection efficiency of the system and

$$\mathcal{L}_\pm = [(\omega_{ex} - \omega_L \pm \Omega_v)^2 + (\kappa_{ex}/2)^2]^{-1} \tag{3}$$

is a Lorentzian field-enhancement profile supported by the exciton resonance $\omega_{ex}$ and exciton damping $\kappa_{ex}$[3,57,58]. The detuning of the laser frequency ($\omega_L$) relative to the exciton resonance determines the direction of energy flow in Raman scattering: red detuning enhances phonon occupation, whereas blue detuning attenuates it. Here we keep the visible power densities below the optomechanical pumping regime, so the $aS/S$ ratio is primarily set by the thermal population. This also further evidenced by the linear increase of both the $S$ and $aS$ intensities with visible powers in the range used here.

MIR pumping drives the system out of thermal equilibrium. MIR-assisted scattering increases the phonon population above its thermal value, and therefore in the presence of a MIR pump the $S$ to $aS$ intensity ratio becomes,

$$\left[\frac{I_{as}}{I_s}\right]_{\mathrm{MIR}} \approx \left(\frac{\omega_L + \Omega_v}{\omega_L - \Omega_v}\right)^4 \left(\frac{\mathcal{L}_-}{\mathcal{L}_+}\right) \zeta \left(\frac{\bar{n}_{\mathrm{MIR}}}{1+\bar{n}_{\mathrm{MIR}}}\right) \tag{4}$$

where $\bar{n}_{\mathrm{MIR}}$ is the phonon population in the presence of the MIR pump.

In MIRAPA, we are interested in the fractional change in phonon population, expressed through the amplification fraction:

$$\Lambda_{\mathrm{MIR}} = (\bar{n}_{\mathrm{MIR}} - \bar{n}_{\mathrm{T}})/\bar{n}_{\mathrm{T}}\ . \tag{5}$$

Direct estimation $\bar{n}_{\mathrm{MIR}}$ from Eq(4) is hindered by contributions from $\varepsilon_{\mathrm{A}}$ and $\varepsilon_{\mathrm{B}}$ excitons and differing electron-phonon coupling strengths across excitonic channels. We therefore take the ratio of intensity ratios with MIR and without MIR, which removes common perfectors and collection efficiencies appearing in Eq(2) and Eq(4). The resulting $\bar{n}_{\mathrm{MIR}}$ under MIRAPA is given (SI2) by:

$$\bar{n}_{\mathrm{MIR}} = \frac{\Re\, \bar{n}_T}{1 - \bar{n}_T(\Re - 1)} \quad (6)$$

where $\Re = \left[\frac{I_{as}}{I_s}\right]_{\mathrm{MIR}} / \left[\frac{I_{as}}{I_s}\right]_0$. This provides a direct estimate of phonon population from experimentally measured $S$ and $aS$ intensities with and without MIR pumping.

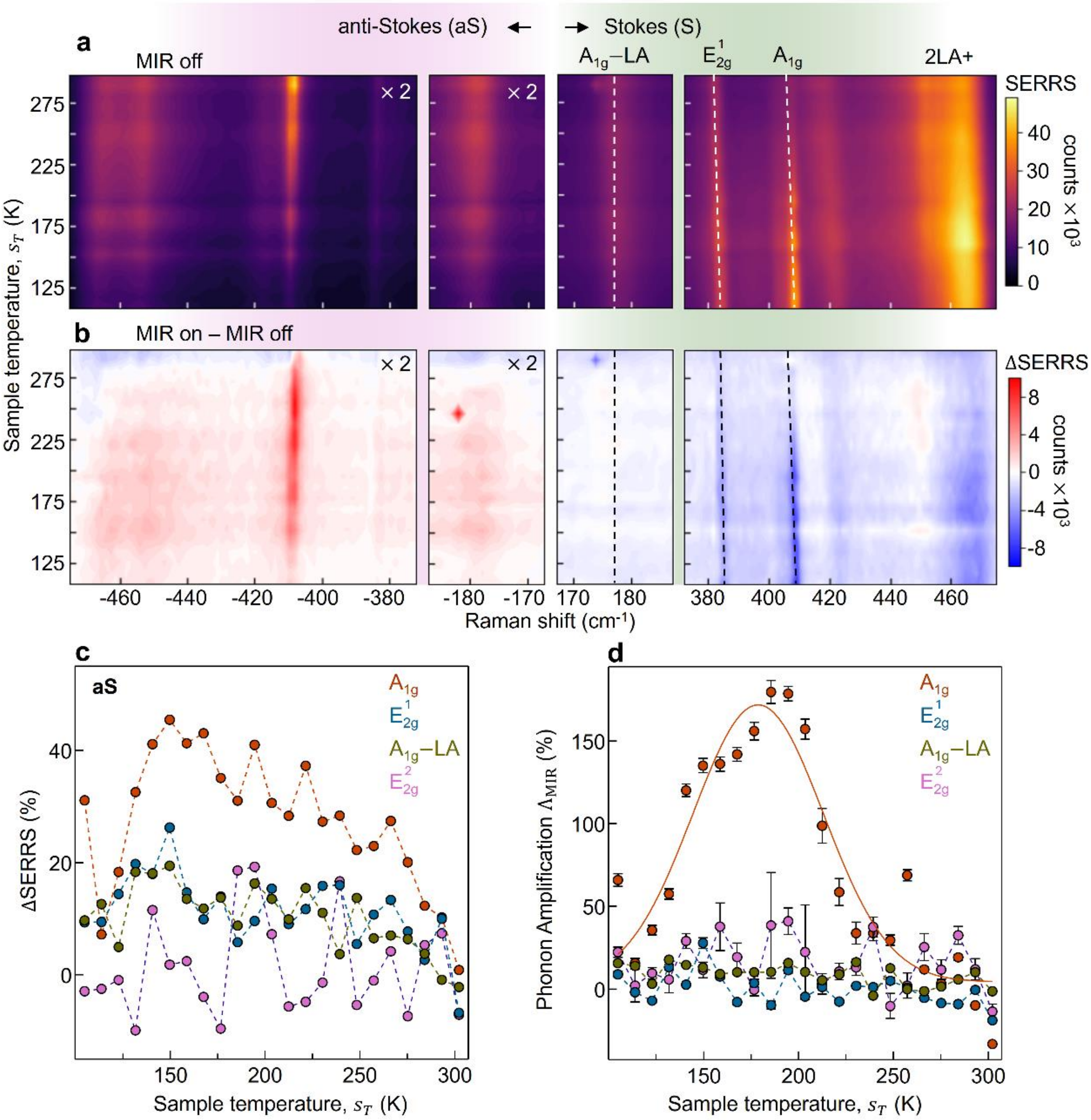

**Figure 2: Temperature-dependent MIR-induced phonon amplification.** (**a**) Stokes and anti-Stokes maps of $MoS_2$ for various sample temperatures without MIR illumination. (**b**) Differential SERRS maps (ΔSERRS) obtained under MIR illumination with power density, $P_{\lambda_{\mathrm{MIR}}} = 3.60\mu\mathrm{W}/\mu\mathrm{m}^2$. (**c**) Percentage change in the anti-SERRS intensity for selected phonon modes as a function of sample temperature. (**d**) MIR-induced phonon amplification from the thermal population estimated by Eq(5) for $\mathrm{A}_{1\mathrm{g}}$, $\mathrm{E}_{2\mathrm{g}}^1$, $\mathrm{A}_{1\mathrm{g}} - \mathrm{LA}$ and $\mathrm{E}_{2\mathrm{g}}^2$ modes, with the phonon population extracted from Eq(6). The solid black curve shows a Gaussian fit to A1g mode, yielding a characteristic temperature width of 54K.

**Temperature-dependent MIRAPA**

Incoherent absorption of visible or MIR photons can lead to heating and suppress selective phonon pumping through optomechanical coupling. To disentangle the roles of exciton resonance and laser heating, we perform temperature-dependent SERRS measurements with and without MIR illumination. The $s_T$ is varied from 105K to 300K in 10K steps (Methods) calibrated using a silicon reference (SI3). At each set temperature we allow sufficient time for thermal equilibrium before acquiring SERRS spectra with and without the MIR pump.

In the absence of the MIR pumping, the $aS$ peak intensity increases with increasing $s_T$. We quantify this by monitoring the peak areas of four phonon modes: $\mathrm{A_{1g}}$, $\mathrm{A_{1g} - LA}$, $\mathrm{E_{2g}^1}$ and $\mathrm{E_{2g}^2}$ (SI4, Fig.S6). A local fitting of these peaks further confirms this, and the extracted relative intensities follow the expected exponential dependence on $1/s_T$ from the Bose-Einstein statistics of the thermal phonon population (SI5, Fig.S9). This confirms that without MIR illumination the phonon population is close to thermal equilibrium. Efficient thermal management is ensured by mounting the flakes on a thin Au foil, which acts as an effective heat sink; Control experiments without the Au support exhibit pronounced thermal defocusing, indicating substantial local heating. We note that while higher visible power densities can optomechanically pump the system out of equilibrium, here the aim is to drive the phonons using MIR photons, whose much lower energy reduces incoherent heating losses.

With MIR pumping at lowered $s_T$ we observe an increase in $aS$ peak intensity for the $\mathrm{A_{1g}}$ phonon with negligible change for the in-plane $\mathrm{E_{2g}^1}$ phonon mode. This is consistent across different flakes that we have measured across different temperatures (SI6, Fig.S10). Plotting the SERRS difference (ΔSERRS, with MIR minus without MIR) as a function of temperature shows that decreasing $s_T$ by 100K from room temperature is sufficient to achieve ~40% amplification of the $\mathrm{A_{1g}}$ mode, whereas other modes ($\mathrm{A_{1g} - LA}$, $\mathrm{E_{2g}^1}$ and $\mathrm{E_{2g}^2}$) show no significant $aS$ intensity change even at 80K (Fig.2c). The multi-phonon 2LA+ feature at 460cm$^{-1}$ shows a modulation that changes sign with temperature (SI7, Fig.S11). At intermediate temperatures we observe a Fano-like lineshape, consistent with a shift effective peak position from higher to lower energy. We attribute this to the sustaining of the van Hove singularity and redistribution among contributions within broad 2LA+ band (e.g., from **K**-point to **M**-point processes), which warrants a dedicated study beyond the scope of this work. The selective amplification of $\mathrm{A_{1g}}$ mode highlights the role of mode symmetry and coupling to electronic excitations.

Estimating the change in phonon population ($\Lambda_{\mathrm{MIR}}$ from Eq(5)) across temperature shows amplification exceeding 150% for $\mathrm{A_{1g}}$ mode with minimal changes in other modes at $s_T$=190K (Fig.2d). The temperature dependence of %$\Lambda_{\mathrm{MIR}}$ is consistent with shifts in the $\varepsilon_{\mathrm{A}}$ and $\varepsilon_{\mathrm{B}}$ exciton resonance frequencies with $s_T$. As temperature increases, the exciton energy approaches the anti-Stokes shifted energy of $\mathrm{A_{1g}}$ scattering, increasing $\Lambda_{\mathrm{MIR}}$ above the base thermal population. At higher $s_T$ the exciton continues to redshift, detuning from the anti-Stokes resonance and reducing $\Lambda_{\mathrm{MIR}}$. Modelling this to a Gaussian to match the exciton lineshape, the width of the $\Lambda_{\mathrm{MIR}}$ curve correlates with the optical damping ($\kappa$=54K). The absence of a comparable response in other modes indicates mode selective pumping despite exciton damping being much larger than the phonon frequency.

We now focus our discussion on $\mathrm{A_{1g}}$ mode and quantify the MIR coupling by estimating the change in the phonon line widths at the peak of $\Lambda_{\mathrm{MIR}}$ in the temperature range of 175-220K. A MIR-induced damping ($\Delta\Gamma_{\mathrm{MIRAPA}}$) is determined as the difference between the phonon linewidths of $\mathrm{A_{1g}}$ mode with ($\Gamma_{\mathrm{MIR}}$) and without MIR ($\Gamma_0$), allowing to estimate the optomechanical coupling rate ($\hbar G$).

$$\Delta\Gamma_{\mathrm{MIRAPA}} = \Gamma_{\mathrm{MIR}} - \Gamma_0 \qquad (7)$$

Across the temperature range of 175K-220K we observe that the $\mathrm{A_{1g}}$ linewidth decreases under MIR, implying $\Delta\Gamma_{\mathrm{MIRAPA}} < 0$. Within the standard dynamical backaction framework of cavity optomechanics, this corresponds to negative optical damping (anti-damping), placing the system in an amplification regime. Assuming an optical damping rate ($\hbar\kappa$) of $4.6$meV from the width of $\Lambda_{\mathrm{MIR}}$ (Fig.2d) and using $\Delta\Gamma_{\mathrm{MIRPA}} \approx 4G^2/\kappa$, we obtain an effective linearised optomechanical coupling $\hbar G$ ~0.07-0.14meV (17-34GHz), which is temperature dependent and placing the system in the weak-coupling regime. Notably, the extracted coupling is orders of magnitude larger than that of typical micro- or nano-mechanical optomechanical systems (kHz–MHz scale), while experiments on THz phonons with 2D materials have reported similar coupling rates[20,21,59] with visible driving frequencies. The observed linewidth narrowing is directly analogous to phonon gain phenomena reported in electrically pumped THz phonon systems. In electrically driven phonon-polariton quantum cascade structures[60], stimulated emission into mixed photon–phonon modes leads to coherent polariton build-up and strong modification of the phonon population, as demonstrated in electrically pumped phonon-polariton lasers. In contrast, our platform operates without electrical injection; insted the MIR field directly modifies the phonon dissipation channel via exciton-mediated optomechanical coupling.

**Comparing Visible and MIR power dependence**

To compare MIRAPA with visible laser-driven pumping of out-of-plane THz phonons, we measured MIR and visible power dependences at a fixed flake location on the flake (Fig.3). As the MIR power density ($P_{\lambda_{\mathrm{MIR}}}$) is increased from 0.36μW/μm$^2$ to 3.6μW/μm$^2$ at fixed visible power density ($P_{\lambda_{\mathrm{L}}}$=280μW/μm$^2$), the $\mathrm{A_{1g}}$ intensity is strongly enhanced accompanied by a small increase in the low frequency $\mathrm{E_{2g}^2}$ mode and in the second order $\mathrm{A_{1g}} - \mathrm{LA}$ and $2\mathrm{LA} +$ peak, while the $\mathrm{E_{2g}^1}$ mode changes minimally (Fig.3a,S12). This is consistent with the mode selectivity observed above, although the experiment here is performed on a different $MoS_2$ flake of similar thickness to before. We note that by using a higher NA and a lower laser power we can observe a strong enhancement through MIRAPA even at room temperature (Fig.S10). By contrast, when visible power is varied in the absence of MIR, a similar increase is observed for the $\mathrm{A_{1g}}$ mode (Fig.3b), while the $\mathrm{E_{2g}^1}$ mode and second order peaks develop a Fano-like distortions for powers >1.1mW/μm$^2$ (black arrow, Fig.3b), consistent with heating as the system exits the optomechanical pumping regime.

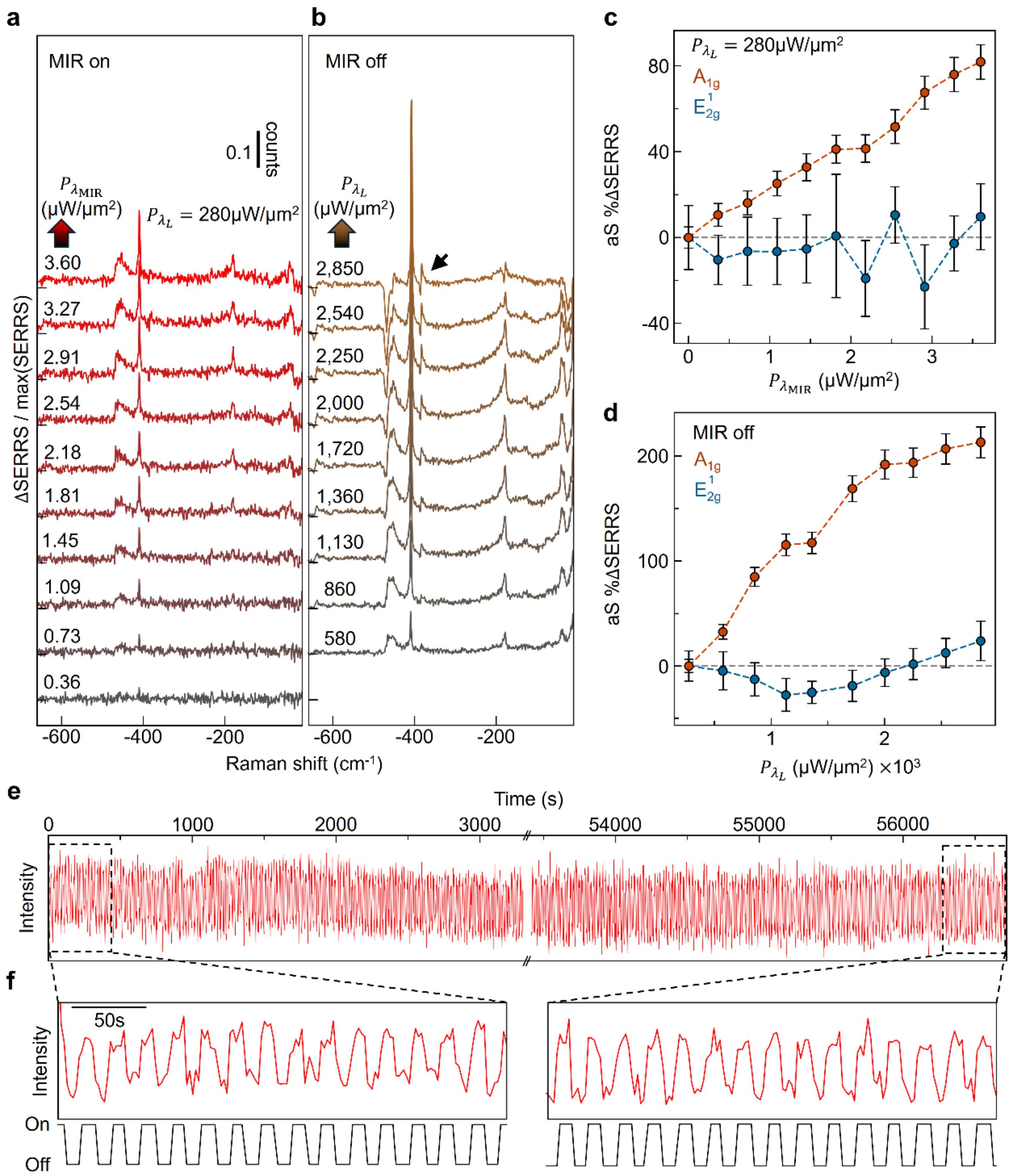

**Figure 3: MIR and visible power dependence of anti-Stokes SERRS.** (**a**) Differential anti-Stokes SERRS spectra for increasing MIR power density $P_{\lambda_{\rm MIR}}$ at fixed visible excitation power $P_{\lambda_{\rm L}}$=280μW/μm². Each spectrum is normalised to the maximum intensity obtained in the absence of MIR illumination, vertically offset for clarity, and labelled with the corresponding MIR power density. (**b**) ΔSERRS $aS$ spectra for increasing $P_{\lambda_{\rm L}}$ with no MIR illumination ($P_{\lambda_{\rm MIR}}$=0). Each spectrum is normalised to its own maximum intensity and labelled with illuminated $P_{\lambda_{\rm L}}$. (**c, d**) The $aS$ percentage change in the ΔSERRS intensity extracted from **a** and **b** respectively for the $\mathrm{A_{1g}}$ and $\mathrm{E^1_{2g}}$ peaks. (**e**) Time-domain modulation of the anti-Stokes intensity under periodic MIR on/off switching over more than 2800 cycles ($P_{\lambda_{\rm MIR}}$ = 3.30μW/μm² and $P_{\lambda_{\rm L}}$=280μW/μm²), demonstrating stable and repeatable MIR-induced modulation.

(**f**) Expanded views of the initial and final 300 s segments of the time trace in (**e**). The black traces show the extracted binary on/off modulation obtained from threshold-based fitting, confirming reproducible MIR-correlated signal changes.

Comparing the $\mathrm{A_{1g}}$ response for MIR and visible power dependences shows >80% change in peak intensity (red circles) in both cases (Fig.3c,d). Critically, the MIR power density required is nearly 300 times less than the visible power density needed for a similar change in the aS amplification. Even after accounting for the visible laser going through the Au foil we find that MIR is 150 times more efficient in phonon pumping (Fig.S13). Thus, MIRAPA selectively prepares the vibrational subsystem by populating specific phonon modes with MIR, while the visible resonant Raman probe provides a sensitive readout of the resulting non-thermal phonon distribution. In the MIR spectral region, the optical susceptibility is dominated by lattice (ionic) contributions rather than electronic transitions, enabling efficient coupling of the MIR field to phonon-related polarizability, particularly for out-of-plane vibrations that experience enhanced local fields near the metal interface. This allows MIR illumination to act as an effective phonon drive with minimal electronic heating.

## MIRPA for MIR upconversion detection

As a final demonstration of stability and sensitivity, we investigate the optical modulation of the anti-Stokes SERRS signal under periodic MIR on/off modulation (Fig.3e,f), using a MIR power density $P_{\lambda_{\mathrm{MIR}}}$=3.30µW/µm$^2$ and visible excitation $P_{\lambda_{\mathrm{L}}}$=280µW/µm$^2$. The anti-Stokes intensity exhibits a clear and reproducible modulation that remains stable over more than 2800 switching cycles. Expanded views of the initial and final 300s segments confirm that the modulation depth and phase remains unchanged throughout the measurement. Binary threshold fitting (black traces in Fig.3f) robustly extracts the MIR-correlated on/off response, ruling out fast artefacts or stochastic fluctuations. The MIR illumination induces a mean modulation depth of approximately 16%, with only ∼1.9% cycle-to-cycle variability (SI9, Fig.S15), demonstrating excellent repeatability over extended time scales.

The long-term stability further enables a quantitative assessment of detector sensitivity. Over the full measurement duration, the residual drift in the anti-Stokes signal is limited to ∼2.9% per hour, consistent with slow processes such as thermal equilibration or gradual coupling evolution rather than noise-dominated or unstable behaviour. Using a background noise level extracted from a featureless spectral region (-800 to -700 cm$^{-1}$) and integrating the anti-Stokes response, we obtain an area-based signal-to-noise ratio exceeding 10$^2$ at a MIR power density of 3.30µW/µm$^2$ applied over a detection area limited by visible laser spot. This corresponds to an effective noise-equivalent power (NEP) 0.27×10$^{-9}$W/√Hz. While this NEP does not yet rival state-of-the-art cryogenically cooled HgCdTe (MCT) detectors which can reach 10$^{-11}$-10$^{-12}$ W/√Hz in the mid-infrared, it is comparable to or competitive with uncooled thermal MCT and bolometric systems operating at room temperature. Importantly, the present scheme achieves this sensitivity through an entirely optical upconversion pathway, without electrical readout, cryogenic cooling, or semiconductor junction engineering[61]. These results motivate further exploration of MIRAPA in 2D excitonic systems integrated with microcavity[62,63] and nanocavity[37,64–66], where strong light-matter coupling and enhanced Purcell factors may push the sensitivity towards the regime of conventional semiconductor photodetectors.

## Conclusion

In conclusion, we establish MIR-assisted phonon amplification (MIRAPA) as a physically distinct and energetically efficient route to coherently bias specific lattice vibrations in a two-dimensional semiconductor. By exploiting the tripartite interaction between the mid-infrared drive, the excitonic resonance, and the out-of-plane mechanical mode, we achieve >80% amplification of the $A_{1g}$ THz phonon in few-layer $MoS_2$. Crucially, the use of MIR excitation reduces parasitic lattice heating by orders of magnitude: achieving comparable phonon enhancement requires ∼300× lower power density than visible pumping. This clearly differentiates MIRAPA from purely photothermal or non-resonant excitation mechanisms and demonstrates that selective vibrational energy injection can be achieved without globally perturbing the crystal lattice. The process remains stable and reproducible over extended operation (>15 hours) and exhibits a noise-equivalent power of ∼0.3nW/√Hz, positioning few-layer $MoS_2$ as an ultrasensitive MIR transducer. Direct frequency matching of the MIR field to the target vibrational mode offers a clear pathway to further increase efficiency while suppressing off-resonant absorption. More broadly, extending MIRAPA to other 2D quantum materials, magnetic van der Waals crystals, or engineered heterostructures could enable deterministic control of individual THz phonons via coupling to the internal resonances of the structures across a wide spectral range[67]. Integration within nanophotonic cavities that co-confine visible and MIR fields would enhance the effective optomechanical coupling rate, opening a realistic route toward stimulated THz phonon emission and ultimately phonon lasing[60]. Collectively, these results position MIR-driven optomechanics as a scalable platform for low-power vibrational control, mid-infrared detection, and engineered light-matter coupling in quantum materials.

## Methods

**Experimental setup.** All Raman spectroscopy measurements are performed with a Renishaw inVia Raman microscope, using a 50x 0.5NA microscope objective to focus the visible light. The spot size for the 633nm laser is 1.7µm$^2$. The light is collected through the same 0.5NA objective with an 1800lines/mm grating and giving a resolution of 1.34cm$^{-1}$. For temperature control a Linkham FTIR600 stage is used with liquid nitrogen cooling from 80K to 300K in 10K increments, with a wait time at each temperature to ensure the sample is thermalised. The temperature is calibrated using a silicon reference (SI3). The windows on the Linkam FTIR600 stage are customised for transparency in the correct optical regime; a glass window for the visible light (incident and scattering) and a zinc selenide (ZnSe) window for the MIR light. A ThorLabs Turnkey QCL is used to provide MIR light at a centre wavelength of 4.65µm. The MIR light is focused using a gold coated parabolic mirror with a reflective focal length of 76.2mm, yielding a spot size of 0.14mm$^2$.

**Sample preparation**

The $MoS_2$ samples are prepared by metal-assisted exfoliation[68,69]. First, a 15nm Au film with a 1nm Ti adhesion layer is deposited onto freshly cleaned fused silica by magnetron sputtering. The Au and Ti targets, from Birmingham Metal Ltd. and Testbourne Ltd., respectively, have purity >99.99%. The base pressure of the Kurt J. Lesker CMS-A UHV deposition system is better than $1\times10^{-8}$ Torr. Shortly after deposition and immediately after exposure to ambient, a freshly cleaved bulk $MoS_2$ crystal from HQ Graphene is gently pressed onto the metal surface and peeled to complete the exfoliation process.

ASSOCIATED CONTENT

**Supporting Information**. A Supporting Information document is also provided, with additional images and information.

AUTHOR INFORMATION

**Corresponding Author**

* Dr Rohit Chikkaraddy, r.chikkaraddy@bham.ac.uk

**Notes**

The authors declare no competing financial interest.

ACKNOWLEDGMENT

We acknowledge funding from UKRI Future Leaders Fellowship, EPSRC (EP/Y008774/1) and Royal Society (RGS/R1/231458). J.Z and F.H acknowledge funding support from EPSRC and SFI (EP/S023321/1).

## REFERENCES

(1) Devereaux, T. P.; Hackl, R. Inelastic Light Scattering from Correlated Electrons. *Rev. Mod. Phys.* **2007**, *79* (1), 175–233. https://doi.org/10.1103/RevModPhys.79.175.

(2) Kampfrath, T.; Tanaka, K.; Nelson, K. A. Resonant and Nonresonant Control over Matter and Light by Intense Terahertz Transients. *Nature Photon* **2013**, *7* (9), 680–690. https://doi.org/10.1038/nphoton.2013.184.

(3) Aspelmeyer, M.; Kippenberg, T. J.; Marquardt, F. Cavity Optomechanics. *Rev. Mod. Phys.* **2014**, *86* (4), 1391–1452. https://doi.org/10.1103/RevModPhys.86.1391.

(4) Roelli, P.; Galland, C.; Piro, N.; Kippenberg, T. J. Molecular Cavity Optomechanics as a Theory of Plasmon-Enhanced Raman Scattering. *Nature Nanotech* **2016**, *11* (2), 164–169. https://doi.org/10.1038/nnano.2015.264.

(5) Mak, K. F.; Shan, J. Photonics and Optoelectronics of 2D Semiconductor Transition Metal Dichalcogenides. *Nature Photon* **2016**, *10* (4), 216–226. https://doi.org/10.1038/nphoton.2015.282.

(6) Wang, Q. H.; Kalantar-Zadeh, K.; Kis, A.; Coleman, J. N.; Strano, M. S. Electronics and Optoelectronics of Two-Dimensional Transition Metal Dichalcogenides. *Nature Nanotech* **2012**, *7* (11), 699–712. https://doi.org/10.1038/nnano.2012.193.

(7) Jariwala, D.; Sangwan, V. K.; Lauhon, L. J.; Marks, T. J.; Hersam, M. C. Emerging Device Applications for Semiconducting Two-Dimensional Transition Metal Dichalcogenides. *ACS Nano* **2014**, *8* (2), 1102–1120. https://doi.org/10.1021/nn500064s.

(8) Basov, D. N.; Averitt, R. D.; Hsieh, D. Towards Properties on Demand in Quantum Materials. *Nature Materials* **2017**, *16* (11), 1077–1088. https://doi.org/10.1038/nmat5017.

(9) Först, M.; Manzoni, C.; Kaiser, S.; Tomioka, Y.; Tokura, Y.; Merlin, R.; Cavalleri, A. Nonlinear Phononics as an Ultrafast Route to Lattice Control. *Nature Physics* **2011**, *7* (11), 854–856. https://doi.org/10.1038/nphys2055.

(10) Sheng, J. Self-Organized Synchronization of Phonon Lasers. *Phys. Rev. Lett.* **2020**, *124* (5). https://doi.org/10.1103/PhysRevLett.124.053604.

(11) Zhang, X.; Qiao, X.-F.; Shi, W.; Wu, J.-B.; Jiang, D.-S.; Tan, P.-H. Phonon and Raman Scattering of Two-Dimensional Transition Metal Dichalcogenides from Monolayer, Multilayer to Bulk Material. *Chem. Soc. Rev.* **2015**, *44* (9), 2757–2785. https://doi.org/10.1039/C4CS00282B.

(12) Wilson, J. A.; Yoffe, A. D. The Transition Metal Dichalcogenides Discussion and Interpretation of the Observed Optical, Electrical and Structural Properties. *Advances in Physics* **1969**, *18* (73), 193–335. https://doi.org/10.1080/00018736900101307.

(13) Molina-Sánchez, A.; Wirtz, L. Phonons in Single-Layer and Few-Layer MoS2 and WS2. *Phys. Rev. B* **2011**, *84* (15), 155413. https://doi.org/10.1103/PhysRevB.84.155413.

(14) Livneh, T.; Spanier, J. E. A Comprehensive Multiphonon Spectral Analysis in MoS2. *2D Mater.* **2015**, *2* (3), 035003. https://doi.org/10.1088/2053-1583/2/3/035003.

(15) Mak, K. F.; Lee, C.; Hone, J.; Shan, J.; Heinz, T. F. Atomically Thin MoS2: A New Direct-Gap Semiconductor. *Phys. Rev. Lett.* **2010**, *105* (13), 136805. https://doi.org/10.1103/PhysRevLett.105.136805.

(16) Carvalho, B. R.; Malard, L. M.; Alves, J. M.; Fantini, C.; Pimenta, M. A. Symmetry-Dependent Exciton-Phonon Coupling in 2D and Bulk MoS2 Observed by Resonance Raman Scattering. *Phys. Rev. Lett.* **2015**, *114* (13), 136403. https://doi.org/10.1103/PhysRevLett.114.136403.

(17) Wang, G.; Chernikov, A.; Glazov, M. M.; Heinz, T. F.; Marie, X.; Amand, T.; Urbaszek, B. Colloquium: Excitons in Atomically Thin Transition Metal Dichalcogenides. *Rev. Mod. Phys.* **2018**, *90* (2), 021001. https://doi.org/10.1103/RevModPhys.90.021001.

(18) Chernikov, A.; Berkelbach, T. C.; Hill, H. M.; Rigosi, A.; Li, Y.; Aslan, B.; Reichman, D. R.; Hybertsen, M. S.; Heinz, T. F. Exciton Binding Energy and Nonhydrogenic Rydberg Series in Monolayer WS2. *Phys. Rev. Lett.* **2014**, *113* (7), 076802. https://doi.org/10.1103/PhysRevLett.113.076802.

(19) Splendiani, A.; Sun, L.; Zhang, Y.; Li, T.; Kim, J.; Chim, C.-Y.; Galli, G.; Wang, F. Emerging Photoluminescence in Monolayer MoS2. *Nano Lett.* **2010**, *10* (4), 1271–1275. https://doi.org/10.1021/nl903868w.
(20) Lai, J.-M.; Sun, Y.-J.; Tan, Q.-H.; Tan, P.-H.; Zhang, J. Laser Cooling of a Lattice Vibration in van Der Waals Semiconductor. *Nano Lett.* **2022**, *22* (17), 7129–7135. https://doi.org/10.1021/acs.nanolett.2c02240.
(21) Zhang, J.; Zhang, Q.; Wang, X.; Kwek, L. C.; Xiong, Q. Resolved-Sideband Raman Cooling of an Optical Phonon in Semiconductor Materials. *Nature Photon* **2016**, *10* (9), 600–605. https://doi.org/10.1038/nphoton.2016.122.
(22) Shlesinger, I.; Vandersmissen, J.; Oksenberg, E.; Verhagen, E.; Koenderink, A. F. Hybrid Cavity-Antenna Architecture for Strong and Tunable Sideband-Selective Molecular Raman Scattering Enhancement. *Science Advances* **2023**, *9* (51), eadj4637. https://doi.org/10.1126/sciadv.adj4637.
(23) Parra-Murillo, C. A.; Santos, M. F.; Monken, C. H.; Jorio, A. Stokes--Anti-Stokes Correlation in the Inelastic Scattering of Light by Matter and Generalization of the Bose-Einstein Population Function. *Phys. Rev. B* **2016**, *93* (12), 125141. https://doi.org/10.1103/PhysRevB.93.125141.
(24) Jorio, A.; Kasperczyk, M.; Clark, N.; Neu, E.; Maletinsky, P.; Vijayaraghavan, A.; Novotny, L. Optical-Phonon Resonances with Saddle-Point Excitons in Twisted-Bilayer Graphene. *Nano Lett.* **2014**, *14* (10), 5687–5692. https://doi.org/10.1021/nl502412g.
(25) Jones, A. M.; Yu, H.; Schaibley, J. R.; Yan, J.; Mandrus, D. G.; Taniguchi, T.; Watanabe, K.; Dery, H.; Yao, W.; Xu, X. Excitonic Luminescence Upconversion in a Two-Dimensional Semiconductor. *Nature Phys* **2016**, *12* (4), 323–327. https://doi.org/10.1038/nphys3604.
(26) Jadczak, J.; Bryja, L.; Kutrowska-Girzycka, J.; Kapuściński, P.; Bieniek, M.; Huang, Y.-S.; Hawrylak, P. Room Temperature Multi-Phonon Upconversion Photoluminescence in Monolayer Semiconductor WS2. *Nat Commun* **2019**, *10* (1), 107. https://doi.org/10.1038/s41467-018-07994-1.
(27) Qi, P.; Dai, Y.; Luo, Y.; Tao, G.; Zheng, L.; Liu, D.; Zhang, T.; Zhou, J.; Shen, B.; Lin, F.; Liu, Z.; Fang, Z. Giant Excitonic Upconverted Emission from Two-Dimensional Semiconductor in Doubly Resonant Plasmonic Nanocavity. *Light Sci Appl* **2022**, *11* (1), 176. https://doi.org/10.1038/s41377-022-00860-2.
(28) Dai, Y.; Qi, P.; Tao, G.; Yao, G.; Shi, B.; Liu, Z.; Liu, Z.; He, X.; Peng, P.; Dang, Z.; Zheng, L.; Zhang, T.; Gong, Y.; Guan, Y.; Liu, K.; Fang, Z. Phonon-Assisted Upconversion in Twisted Two-Dimensional Semiconductors. *Light Sci Appl* **2023**, *12* (1), 6. https://doi.org/10.1038/s41377-022-01051-9.
(29) Lien, D.-H.; Uddin, S. Z.; Yeh, M.; Amani, M.; Kim, H.; Ager, J. W.; Yablonovitch, E.; Javey, A. Electrical Suppression of All Nonradiative Recombination Pathways in Monolayer Semiconductors. *Science* **2019**, *364* (6439), 468–471. https://doi.org/10.1126/science.aaw8053.
(30) Queisser, H. J.; Haller, E. E. Defects in Semiconductors: Some Fatal, Some Vital. *Science* **1998**, *281* (5379), 945–950. https://doi.org/10.1126/science.281.5379.945.
(31) Mignuzzi, S.; Pollard, A. J.; Bonini, N.; Brennan, B.; Gilmore, I. S.; Pimenta, M. A.; Richards, D.; Roy, D. Effect of Disorder on Raman Scattering of Single-Layer MoS2. *Phys. Rev. B* **2015**, *91* (19), 195411. https://doi.org/10.1103/PhysRevB.91.195411.
(32) Zhu, Y.; Zhang, Z.; Wang, Y.; Sarkar, S.; Li, Y.; Yan, H.; Ishibe-Veiga, L.; Bagri, A.; Yang, Z. J.; Ramsden, H.; Eda, G.; Hoye, R. L. Z.; Wang, Y.; Chhowalla, M. Effect of Substrate on Sulfur Vacancy Defect-Mediated Photoluminescence in Two-Dimensional MoS2. *J. Phys. Chem. C* **2025**, *129* (17), 8294–8302. https://doi.org/10.1021/acs.jpcc.4c08491.
(33) Yang, Z. J.; Li, Z.; Loh, L.; Moloney, J.; Walmsley, J.; Li, J.; Chen, Y.; Liu, L.; Zang, H.; Yan, H.; Sarkar, S.; Day, J.; Wang, Y.; Chhowalla, M. Scalable Manufacture of Nearly Pure-Phase Metallic MoS2 Nanosheets. *Nat. Mater.* **2026**, 1–7. https://doi.org/10.1038/s41563-026-02480-2.

(34) Yin, J.; Zhang, M.; Tan, Y.; Guo, Z.; He, H.; Lan, L.; Cheng, J.-X. Video-Rate Mid-Infrared Photothermal Imaging by Single-Pulse Photothermal Detection per Pixel. *Science Advances* **2023**, *9* (24), eadg8814. https://doi.org/10.1126/sciadv.adg8814.

(35) Cheng, J.-X.; Xie, X. S. Vibrational Spectroscopic Imaging of Living Systems: An Emerging Platform for Biology and Medicine. *Science* **2015**, *350* (6264), aaa8870. https://doi.org/10.1126/science.aaa8870.

(36) Xu, T.; Zhong, F.; Wang, P.; Wang, Z.; Ge, X.; Wang, J.; Wang, H.; Zhang, K.; Zhang, Z.; Zhao, T.; Yu, Y.; Luo, M.; Wang, Y.; Jiang, R.; Wang, F.; Chen, F.; Liu, Q.; Hu, W. Van Der Waals Mid-Wavelength Infrared Detector Linear Array for Room Temperature Passive Imaging. *Science Advances* **2024**, *10* (31), eadn0560. https://doi.org/10.1126/sciadv.adn0560.

(37) Wang, H.; Lee, D.; Cao, Y.; Bi, X.; Du, J.; Miao, K.; Wei, L. Bond-Selective Fluorescence Imaging with Single-Molecule Sensitivity. *Nat. Photon.* **2023**, *17* (10), 846–855. https://doi.org/10.1038/s41566-023-01243-8.

(38) Calvin, A.; Eierman, S.; Peng, Z.; Brzeczek, M.; Satterthwaite, L.; Patterson, D. Single Molecule Infrared Spectroscopy in the Gas Phase. *Nature* **2023**, *621* (7978), 295–299. https://doi.org/10.1038/s41586-023-06351-7.

(39) Rustamkulov, Z.; Sing, D. K.; Mukherjee, S.; May, E. M.; Kirk, J.; Schlawin, E.; Line, M. R.; Piaulet, C.; Carter, A. L.; Batalha, N. E.; Goyal, J. M.; López-Morales, M.; Lothringer, J. D.; MacDonald, R. J.; Moran, S. E.; Stevenson, K. B.; Wakeford, H. R.; Espinoza, N.; Bean, J. L.; Batalha, N. M.; Benneke, B.; Berta-Thompson, Z. K.; Crossfield, I. J. M.; Gao, P.; Kreidberg, L.; Powell, D. K.; Cubillos, P. E.; Gibson, N. P.; Leconte, J.; Molaverdikhani, K.; Nikolov, N. K.; Parmentier, V.; Roy, P.; Taylor, J.; Turner, J. D.; Wheatley, P. J.; Aggarwal, K.; Ahrer, E.; Alam, M. K.; Alderson, L.; Allen, N. H.; Banerjee, A.; Barat, S.; Barrado, D.; Barstow, J. K.; Bell, T. J.; Blecic, J.; Brande, J.; Casewell, S.; Changeat, Q.; Chubb, K. L.; Crouzet, N.; Daylan, T.; Decin, L.; Désert, J.; Mikal-Evans, T.; Feinstein, A. D.; Flagg, L.; Fortney, J. J.; Harrington, J.; Heng, K.; Hong, Y.; Hu, R.; Iro, N.; Kataria, T.; Kempton, E. M.-R.; Krick, J.; Lendl, M.; Lillo-Box, J.; Louca, A.; Lustig-Yaeger, J.; Mancini, L.; Mansfield, M.; Mayne, N. J.; Miguel, Y.; Morello, G.; Ohno, K.; Palle, E.; Petit dit de la Roche, D. J. M.; Rackham, B. V.; Radica, M.; Ramos-Rosado, L.; Redfield, S.; Rogers, L. K.; Shkolnik, E. L.; Southworth, J.; Teske, J.; Tremblin, P.; Tucker, G. S.; Venot, O.; Waalkes, W. C.; Welbanks, L.; Zhang, X.; Zieba, S. Early Release Science of the Exoplanet WASP-39b with JWST NIRSpec PRISM. *Nature* **2023**, *614* (7949), 659–663. https://doi.org/10.1038/s41586-022-05677-y.

(40) Lee, J.-U.; Park, J.; Son, Y.-W.; Cheong, H. Anomalous Excitonic Resonance Raman Effects in Few-Layered MoS2. *Nanoscale* **2015**, *7* (7), 3229–3236. https://doi.org/10.1039/C4NR05785F.

(41) Lee, C.; Yan, H.; Brus, L. E.; Heinz, T. F.; Hone, J.; Ryu, S. Anomalous Lattice Vibrations of Single- and Few-Layer MoS2. *ACS Nano* **2010**, *4* (5), 2695–2700. https://doi.org/10.1021/nn1003937.

(42) Chakraborty, B.; Matte, H. S. S. R.; Sood, A. K.; Rao, C. N. R. Layer-Dependent Resonant Raman Scattering of a Few Layer MoS2. *Journal of Raman Spectroscopy* **2013**, *44* (1), 92–96. https://doi.org/10.1002/jrs.4147.

(43) Windom, B. C.; Sawyer, W. G.; Hahn, D. W. A Raman Spectroscopic Study of MoS2 and MoO3: Applications to Tribological Systems. *Tribol Lett* **2011**, *42* (3), 301–310. https://doi.org/10.1007/s11249-011-9774-x.

(44) Frey, G. L.; Tenne, R.; Matthews, M. J.; Dresselhaus, M. S.; Dresselhaus, G. Raman and Resonance Raman Investigation of MoS2 Nanoparticles. *Phys. Rev. B* **1999**, *60* (4), 2883–2892. https://doi.org/10.1103/PhysRevB.60.2883.

(45) Gontijo, R. N.; Gadelha, A.; Silveira, O. J.; Nunes, R. W.; Pimenta, M. A.; Righi, A.; Fantini, C. Probing Combinations of Acoustic Phonons in MoS2 by Intervalley Double-Resonance Raman Scattering. *Phys. Rev. B* **2021**, *103* (4), 045411. https://doi.org/10.1103/PhysRevB.103.045411.

(46) Gołasa, K.; Grzeszczyk, M.; Bożek, R.; Leszczyński, P.; Wysmołek, A.; Potemski, M.; Babiński, A. Resonant Raman Scattering in MoS2—From Bulk to Monolayer. *Solid State Communications* **2014**, *197*, 53–56. https://doi.org/10.1016/j.ssc.2014.08.009.

(47) Gontijo, R. N.; Resende, G. C.; Fantini, C.; Carvalho, B. R. Double Resonance Raman Scattering Process in 2D Materials. *Journal of Materials Research* **2019**, *34* (12), 1976–1992. https://doi.org/10.1557/jmr.2019.167.
(48) Carvalho, B. R.; Wang, Y.; Mignuzzi, S.; Roy, D.; Terrones, M.; Fantini, C.; Crespi, V. H.; Malard, L. M.; Pimenta, M. A. Intervalley Scattering by Acoustic Phonons in Two-Dimensional MoS2 Revealed by Double-Resonance Raman Spectroscopy. *Nat Commun* **2017**, *8* (1), 14670. https://doi.org/10.1038/ncomms14670.
(49) Gołasa, K.; Grzeszczyk, M.; Leszczyński, P.; Faugeras, C.; Nicolet, A. A. L.; Wysmołek, A.; Potemski, M.; Babiński, A. Multiphonon Resonant Raman Scattering in MoS2. *Applied Physics Letters* **2014**, *104* (9), 092106. https://doi.org/10.1063/1.4867502.
(50) Trovatello, C.; Miranda, H. P. C.; Molina-Sánchez, A.; Borrego-Varillas, R.; Manzoni, C.; Moretti, L.; Ganzer, L.; Maiuri, M.; Wang, J.; Dumcenco, D.; Kis, A.; Wirtz, L.; Marini, A.; Soavi, G.; Ferrari, A. C.; Cerullo, G.; Sangalli, D.; Conte, S. D. Strongly Coupled Coherent Phonons in Single-Layer MoS2. *ACS Nano* **2020**, *14* (5), 5700–5710. https://doi.org/10.1021/acsnano.0c00309.
(51) Mertens, J.; Shi, Y.; Molina-Sánchez, A.; Wirtz, L.; Yang, H. Y.; Baumberg, J. J. Excitons in a Mirror: Formation of "Optical Bilayers" Using MoS2 Monolayers on Gold Substrates. *Appl. Phys. Lett.* **2014**, *104* (19), 191105. https://doi.org/10.1063/1.4876475.
(52) Kasperczyk, M.; Jorio, A.; Neu, E.; Maletinsky, P.; Novotny, L. Stokes–Anti-Stokes Correlations in Diamond. *Opt. Lett., OL* **2015**, *40* (10), 2393–2396. https://doi.org/10.1364/OL.40.002393.
(53) Freitas, T. A.; Machado, P.; Valente, L.; Sier, D.; Corrêa, R.; Saito, R.; Galland, C.; Santos, M. F.; Monken, C. H.; Jorio, A. Microscopic Origin of Polarization-Entangled Stokes--Anti-Stokes Photons in Diamond. *Phys. Rev. A* **2023**, *108* (5), L051501. https://doi.org/10.1103/PhysRevA.108.L051501.
(54) Tschannen, C. D.; Frimmer, M.; Vasconcelos, T. L.; Shi, L.; Pichler, T.; Novotny, L. Tip-Enhanced Stokes–Anti-Stokes Scattering from Carbyne. *Nano Lett.* **2022**, *22* (8), 3260–3265. https://doi.org/10.1021/acs.nanolett.2c00154.
(55) Vento, V.; Tarrago Velez, S.; Pogrebna, A.; Galland, C. Measurement-Induced Collective Vibrational Quantum Coherence under Spontaneous Raman Scattering in a Liquid. *Nat Commun* **2023**, *14* (1), 2818. https://doi.org/10.1038/s41467-023-38483-9.
(56) Khurgin, J. B. Feasibility of Resonant Raman Cooling and Radiation Balanced Lasing in Semiconductors. *J. Opt. Soc. Am. B, JOSAB* **2022**, *39* (1), 338–344. https://doi.org/10.1364/JOSAB.447676.
(57) Esteban, R.; Baumberg, J. J.; Aizpurua, J. Molecular Optomechanics Approach to Surface-Enhanced Raman Scattering. *Acc. Chem. Res.* **2022**, *55* (14), 1889–1899. https://doi.org/10.1021/acs.accounts.1c00759.
(58) Benz, F.; Schmidt, M. K.; Dreismann, A.; Chikkaraddy, R.; Zhang, Y.; Demetriadou, A.; Carnegie, C.; Ohadi, H.; de Nijs, B.; Esteban, R.; Aizpurua, J.; Baumberg, J. J. Single-Molecule Optomechanics in "Picocavities." *Science* **2016**, *354* (6313), 726–729. https://doi.org/10.1126/science.aah5243.
(59) Xu, Y.; Hu, H.; Chen, W.; Suo, P.; Zhang, Y.; Zhang, S.; Xu, H. Phononic Cavity Optomechanics of Atomically Thin Crystal in Plasmonic Nanocavity. *ACS Nano* **2022**, *16* (8), 12711–12719. https://doi.org/10.1021/acsnano.2c04478.
(60) Ohtani, K.; Meng, B.; Franckié, M.; Bosco, L.; Ndebeka-Bandou, C.; Beck, M.; Faist, J. An Electrically Pumped Phonon-Polariton Laser. *Science Advances* **2019**, *5* (7), eaau1632. https://doi.org/10.1126/sciadv.aau1632.
(61) Koppens, F. H. L.; Mueller, T.; Avouris, P.; Ferrari, A. C.; Vitiello, M. S.; Polini, M. Photodetectors Based on Graphene, Other Two-Dimensional Materials and Hybrid Systems. *Nature Nanotech* **2014**, *9* (10), 780–793. https://doi.org/10.1038/nnano.2014.215.
(62) Dufferwiel, S.; Schwarz, S.; Withers, F.; Trichet, A. a. P.; Li, F.; Sich, M.; Del Pozo-Zamudio, O.; Clark, C.; Nalitov, A.; Solnyshkov, D. D.; Malpuech, G.; Novoselov, K. S.; Smith, J. M.; Skolnick, M. S.; Krizhanovskii, D. N.; Tartakovskii, A. I. Exciton–Polaritons in van Der Waals Heterostructures

Embedded in Tunable Microcavities. *Nat Commun* **2015**, *6* (1), 8579. https://doi.org/10.1038/ncomms9579.
(63) Liu, X.; Galfsky, T.; Sun, Z.; Xia, F.; Lin, E.; Lee, Y.-H.; Kéna-Cohen, S.; Menon, V. M. Strong Light–Matter Coupling in Two-Dimensional Atomic Crystals. *Nat Photon* **2015**, *9* (1), 30–34. https://doi.org/10.1038/nphoton.2014.304.
(64) Kleemann, M.-E.; Chikkaraddy, R.; Alexeev, E. M.; Kos, D.; Carnegie, C.; Deacon, W.; de Pury, A. C.; Große, C.; de Nijs, B.; Mertens, J.; Tartakovskii, A. I.; Baumberg, J. J. Strong-Coupling of WSe 2 in Ultra-Compact Plasmonic Nanocavities at Room Temperature. *Nature Communications* **2017**, *8* (1), 1296. https://doi.org/10.1038/s41467-017-01398-3.
(65) Quan, J.; Cotrufo, M.; Chand, S.; Jiang, X.; Liu, Z.; Mejia, E.; Wang, W.; Taniguchi, T.; Watanabe, K.; Grosso, G.; Li, X.; Alù, A. On-Site Enhancement and Control of Spin-Forbidden Dark Excitons in a Plasmonic Heterostructure. *Nat. Photon.* **2026**, *20* (1), 49–54. https://doi.org/10.1038/s41566-025-01788-w.
(66) Neuman, T. Quantum Description of Surface-Enhanced Resonant Raman Scattering within a Hybrid-Optomechanical Model. *Phys. Rev. A* **2019**, *100* (4). https://doi.org/10.1103/PhysRevA.100.043422.
(67) Du, L.; Molas, M. R.; Huang, Z.; Zhang, G.; Wang, F.; Sun, Z. Moiré Photonics and Optoelectronics. *Science* **2023**, *379* (6639), eadg0014. https://doi.org/10.1126/science.adg0014.
(68) Velický, M.; Donnelly, G. E.; Hendren, W. R.; McFarland, S.; Scullion, D.; DeBenedetti, W. J. I.; Correa, G. C.; Han, Y.; Wain, A. J.; Hines, M. A.; Muller, D. A.; Novoselov, K. S.; Abruña, H. D.; Bowman, R. M.; Santos, E. J. G.; Huang, F. Mechanism of Gold-Assisted Exfoliation of Centimeter-Sized Transition-Metal Dichalcogenide Monolayers. *ACS Nano* **2018**, *12* (10), 10463–10472. https://doi.org/10.1021/acsnano.8b06101.
(69) Ziewer, J.; Ghosh, A.; Hanušová, M.; Pirker, L.; Frank, O.; Velický, M.; Grüning, M.; Huang, F. Strain-Induced Decoupling Drives Gold-Assisted Exfoliation of Large-Area Monolayer 2D Crystals. *Advanced Materials* **2025**, *37* (14), 2419184. https://doi.org/10.1002/adma.202419184.